\documentclass[fleqn,usenatbib,a4paper]{mnras}

\usepackage{times}
\usepackage{flushend}
\usepackage{graphicx}

\newcommand{\Msun}{\ensuremath{\,{\rm M}_\odot}}                  % Solar mass symbol
\newcommand{\Rsun}{\ensuremath{\,{\rm R}_\odot}}                  % Solar radius symbol
\newcommand{\Teff}{\ensuremath{T_{\rm eff}}}                      % Effective temperature symbol
\newcommand{\Mjup}{\ensuremath{\,{\rm M}_{\rm Jup}}}              % Jupiter mass symbol
\newcommand{\Rjup}{\ensuremath{\,{\rm R}_{\rm Jup}}}              % Jupiter radius symbol
\newcommand{\Teq}{\ensuremath{T_{\rm eq}^{\,\prime}}}             % Equilibrium temperature symbol
\newcommand{\safronov}{\ensuremath{\Theta}}                       % Safronov number symbol
\newcommand{\Porb}{\ensuremath{P_{\rm orb}}}                      % Orbital period symbol
\newcommand{\kms}{\,km\,s$^{-1}$}                                 % km/s symbol
\newcommand{\ms}{\,m\,s$^{-1}$}                                   % m/s symbol
\newcommand{\mss}{\,m\,s$^{-2}$}                                  % m/s^2 symbol
\newcommand{\as}{\ensuremath{^{\prime\prime}}}                    % Arcsecond symbol
\newcommand{\am}{\ensuremath{^\prime}}                            % Arcminute symbol
\newcommand{\pjup}{\ensuremath{\,\rho_{\rm Jup}}}                 % Jupiter density symbol
\newcommand{\psun}{\ensuremath{\,\rho_\odot}}                     % Solar density symbol
\newcommand{\chir}{\ensuremath{\chi_\nu^{\,2}}}                   % Reduced chi-squared symbol
\newcommand{\mc}[1]{\multicolumn{2}{c}{#1}}

\newcommand{\reff}[1]{{#1}}

\title[Transit timings for the HATS-18 system]
      {A search for transit timing variations in the HATS-18 planetary system\thanks{Based on data collected by MiNDSTEp with the Danish 1.54\,m telescope at the ESO La Silla Observatory.}}

\author[Southworth et al.]
       {John Southworth$^{1}$, A.\ J.\ Barker$^{2}$, T.\ C.\ Hinse$^{3,4}$, Y.\ Jongen$^{5}$, M.\ Dominik$^{6}$, \newauthor
        U.\ G.\ J{\o}rgensen$^{7}$, P.\ Longa-Pe{\~n}a$^{8}$, S.\ Sajadian$^{9}$, C.\ Snodgrass$^{10}$, J.\ Tregloan-Reed$^{11}$, \newauthor
        N.\ Bach-M{\o}ller$^{7,12}$, M.\ Bonavita$^{10}$, V.\ Bozza$^{13,14}$, M.\ J.\ Burgdorf$^{15}$, R.\ {Figuera Jaimes}$^{16}$, \newauthor
        Ch.\ Helling$^{12,17}$, J.\ A.\ Hitchcock$^{6}$, M.\ Hundertmark$^{18}$, E.\ Khalouei$^{19}$, H.\ Korhonen$^{20}$, \newauthor
        L.\ Mancini$^{21,22,23,24}$, N.\ Peixinho$^{25}$, S.\ Rahvar$^{26}$, M.\ Rabus$^{27}$, J.\ Skottfelt$^{28}$, P.\ Spyratos$^{1}$
        \\
        $^{1}$\,Astrophysics Group, Keele University, Staffordshire, ST5 5BG, UK \\
        $^{2}$\,Department of Applied Mathematics, School of Mathematics, University of Leeds, Leeds, LS2 9JT, UK \\
        $^{3}$\,Institute of Astronomy, Faculty of Physics, Astronomy and Informatics, Nicolaus Copernicus University, ul. Grudziadzka 5, 87-100 Toru{\'n}, Poland \\
        $^{4}$\,Chungnam National University, Department of Astronomy, Space Science and Geology, Daejeon, South Korea \\
        $^{5}$\,Observatoire de Vaison-La-Romaine, D\'epartementale 51, pr\`es du Centre Equestre au Palis -- 84110 Vaison-La-Romaine, France \\
        $^{6}$\,Centre for Exoplanet Science, SUPA, School of Physics \& Astronomy, University of St Andrews, North Haugh, St Andrews KY16 9SS, UK \\
        $^{7}$\,Centre for ExoLife Sciences, Niels Bohr Institute, University of Copenhagen, {\O}ster Voldgade 5, 1350 Copenhagen, Denmark \\
        $^{8}$\,Centro de Astronom\'{\i}a, CITEVA, Universidad de Antofagasta, Av.\ Angamos 601, Antofagasta, Chile \\
        $^{9}$\,Department of Physics, Isfahan University of Technology, Isfahan 84156-83111, Iran \\
        $^{10}$\,Institute for Astronomy, University of Edinburgh, Royal Observatory, Edinburgh EH9 3HJ, UK \\
        $^{11}$\,Instituto de Astronomia y Ciencias Planetarias de Atacama,  Universidad de Atacama, Copayapu 485,  Copiapo, Chile \\
        $^{12}$\,Space Research Institute, Austrian Academy of Sciences, Schmiedlstrasse 6, 8042 Graz, Austria \\
        $^{13}$\,Dipartimento di Fisica ``E.R. Caianiello'', Universit{\`a} di Salerno, Via Giovanni Paolo II 132, 84084, Fisciano, Italy \\
        $^{14}$\,Istituto Nazionale di Fisica Nucleare, Sezione di Napoli, Napoli, Italy \\
        $^{15}$\,Universit{\"a}t Hamburg, Faculty of Mathematics, Informatics and Natural Sciences, Bundesstra\ss{}e 55, 20146 Hamburg, Germany \\
        $^{16}$\,Facultad de Ingenier\'{\i}a y Tecnolog\'{\i}a, Universidad San Sebastian, General Lagos 1163, Valdivia 5110693, Chile \\
        $^{17}$\,TU Graz, Fakult\"at f\"ur Mathematik, Physik und Geod\"asie, Petersgasse 16, 8010 Graz, Austria \\
        $^{18}$\,Astronomisches Rechen-Institut, Zentrum f{\"u}r Astronomie der Universit{\"a}t Heidelberg (ZAH), 69120 Heidelberg, Germany \\
        $^{19}$\,Astronomy Research Center, Research Institute of Basic Sciences, Seoul National University, 1 Gwanak-ro, Gwanak-gu, Seoul 08826, Korea \\
        $^{20}$\,European Southern Observatory (ESO), Alonso de C\'ordova 3107, Vitacura, Santiago, Chile \\
        $^{21}$\,Department of Physics, University of Rome ``Tor Vergata'', Via della Ricerca Scientifica 1, I-00133, Rome, Italy \\
        $^{22}$\,Max Planck Institute for Astronomy, Königstuhl 17, D-69117, Heidelberg, Germany \\
        $^{23}$\,INAF –- Turin Astrophysical Observatory, via Osservatorio 20, I-10025, Pino Torinese, Italy \\
        $^{24}$\,International Institute for Advanced Scientific Studies (IIASS), Via G.\ Pellegrino 19, I-84019, Vietri sul Mare (SA), Italy \\
        $^{25}$\,Instituto de Astrof\'{\i}sica e Ci\^{e}ncias do Espa\c{c}o, Universidade de Coimbra, PT3040-004 Coimbra, Portugal \\
        $^{26}$\,Department of Physics, Sharif University of Technology, PO Box 11155-9161 Tehran, Iran \\
        $^{27}$\,Departamento de Matem\'atica y F\'{\i}sica Aplicadas, Facultad de Ingenier\'{\i}a, Universidad Cat\'olica de la Sant\'{\i}sima Concepci\'on, Alonso de Rivera 2850, Concepci\'on, Chile \\
        $^{28}$\,Centre for Electronic Imaging, Department of Physical Sciences, The Open University, Milton Keynes, MK7 6AA, UK
        }

\begin{document} \maketitle %%%%%%%%%%%%%%%%%%%%%%%%%%%%%%%%%%%%%%%%%%%%%%%%%%%%%%%%%%%%%%%%%%%%%%%%%%%%%%%%%%%%%%%%%%%%%%%%%%%%%%%%%%%%%%%%%%%%%%%%%
%%%%%%%%%%%%%%%%%%%%%%%%%%%%%%%%%%%%%%%%%%%%%%%%%%%%%%%%%%%%%%%%%%%%%%%%%%%%%%%%%%%%%%%%%%%%%%%%%%%%%%%%%%%%%%%%%%%%%%%%%%%%%%%%%%%%%%%%%%%%%%%%%%%%%

% \clearpage

\begin{abstract}
HATS-18\,b is a transiting planet with a large mass and a short orbital period, and is one of the best candidates for the detection of orbital decay induced by tidal effects. We present extensive photometry of HATS-18 from which we measure 27 times of mid-transit. Two further transit times were measured from data from the Transiting Exoplanet Survey Satellite (TESS) \reff{and three more taken from the literature}. The transit timings were fitted with linear and quadratic ephemerides and an upper limit on orbital decay was determined. This corresponds to a lower limit on the modified stellar tidal quality factor of $Q_\star^{\,\prime} > 10^{5.11 \pm 0.04}$. This is at the cusp of constraining the presence of enhanced tidal dissipation due to internal gravity waves. We also refine the measured physical properties of the HATS-18 system, place upper limits on the masses of third bodies, and compare the relative performance of TESS and the 1.54\,m Danish Telescope in measuring transit times for this system.
\end{abstract}


%%%%%%%%%%%%%%%%%%%%%%%%%%%%%%%%%%%%%%%%%%%%%%%%%%%%%%%%%%%%%%%%%%%%%%%%%%%%%%%%%%%%%%%%%%%%%%%%%%%%%%%%%%%%%%%%%%%%%%%%%%%%%%%%%%%%%%%%%%%%%%%%%%%%%

\section{Introduction}
\label{sec:intro}

The hot Jupiters are a class of extrasolar planet characterised by a relatively large mass and a short orbital period. As a result, they are expected to undergo strong tidal interactions with their host stars \citep[e.g.][]{Ogilvie14araa,Mathis19conf}. This will lead to exchanges of angular momentum and energy between the stellar rotation and the planetary orbit. Since the host stars of most hot Jupiters have a rotational frequency below that of their planet's \reff{orbital frequency}, one manifestation of this interaction is orbital decay: a decrease in the orbital period of the planet and therefore the progressively earlier occurrence of transits in a system. Hence, observations of transit times can constrain the efficiency of such tidal interactions \citep[e.g.][]{Birkby+14mn,Maciejewski+18aca,Patra+20aj}. This in turn is relevant as a window on stellar structure \citep{Love11book} and on the ultimate fate of short-period planets \citep{Levrard++09apj,Jackson++09apj,Matsumura+10apj}.

It is therefore helpful to search for transit time variations (TTVs) in systems containing hot Jupiters. Because the effects of tides are small on a human timescale, it is wise to concentrate this search on the most promising systems. \citet{Maciejewski+18aca} used a result from \citet{GoldreichSoter66icar} to develop an equation giving the expected shift in transit times as a function of the orbital period and mass ratio of a system plus the tidal efficiency and size of the host star \reff{(their eq.\,2)}. We used this approach to rank all transiting extrasolar planets (TEPs) in TEPCat\footnote{The Transiting Extrasolar Planet Catalogue \citep{Me11mn} is at: \texttt{https://www.astro.keele.ac.uk/jkt/tepcat/}} in decreasing order of predicted transit shift over a ten-year period. The planetary system HATS-18 was found to be one of the best candidates, so we embarked on a campaign to determine precise transit times for this system over a period of several years.

HATS-18 was observed in the course of the HATSouth transit survey \citep{Bakos+13pasp} and its detection and characterisation was announced by \citet{Penev+16aj}. It consists of a \reff{solar-type} star (1.0\Msun, 1.0\Rsun) with a gas giant (2.0\Mjup, 1.3\Rjup) in an orbit with a very short period (0.838\,d). The properties of the system were measured \citep{Penev+16aj} based primarily on a single high-quality transit light curve, six high-precision radial velocity (RV) measurements, and \reff{predictions from the} Yonsei-Yale theoretical stellar evolutionary models \citep{Demarque+04apjs}. \citet{Penev+16aj} also presented a detailed analysis of the tidal characteristics of the system, highlighting its importance in constraining tidal theory.

There has been little subsequent (published) work on the nature of HATS-18. \citet{Patra+20aj} presented two times of minimum obtained using a 60\,cm telescope which were consistent with a constant orbital period. The system has also been observed in two sectors by the NASA Transiting Exoplanet Survey Satellite (TESS; \citealt{Ricker+15jatis}). We used both datasets in our own analysis in order to obtain the best constraints on the orbital evolution of HATS-18.

Since our work on HATS-18 began, several theoretical studies of tidal dissipation have highlighted the importance of internal gravity waves in hot Jupiter systems and selected HATS-18 as a promising candidate \citep{Barker20mn,MaFuller21apj}. Internal gravity waves are likely to be the dominant tidal mechanism in slowly rotating solar-type main-sequence stars with radiative cores \citep[e.g.][]{BarkerOgilvie10mn,Chernov++17mn,Barker20mn,MaFuller21apj}. In particular, the star HATS-18\,A is one in which tidally-excited gravity waves are marginally predicted to break in the stellar core, depending strongly on the age of the star. When wave breaking occurs, stellar tidal dissipation is predicted to be efficient and can cause planetary orbital decay on Myr timescales for planets with orbital periods of approximately one day \citep[e.g.][]{BarkerOgilvie10mn}. Such orbital decay is potentially observable in HATS-18, thus supporting our aim to search for TTVs in this system.

It is important to remember that there are multiple possible causes of TTVs. Orbital decay has been confidently detected in only one system, WASP-12 \citep{Hebb+09apj,Maciejewski+16aa,Patra+17aj,Maciejewski+18aca}. WASP-4 is also known to have a decreasing orbital period \citep{Bouma+19aj,Me+19mn} but the origin of this is not yet settled \citep{Baluev+20mn,Bouma+20apj,Turner+22aj}. TTVs can arise due to the light-time effect caused by long-period companions in a system, as has been found for TrES-5 \citep{Maciejewski+21aa}, WASP-148 \citep{Maciejewski+20aca} and HAT-P-26 \citep{Vonessen+19aa}. TTVs can also be caused by apsidal precession \citep{Patra+17aj}, starspots \citep{WatsonDhillon04mn,Oshagh+13aa}, the Applegate mechanism \citep{Applegate92apj,WatsonMarsh10mn} and gravitational perturbations in multi-planetary systems \citep[e.g.][]{HolmanMurray05sci,Agol+05mn,Rowe+14apj}. \reff{Some of these effects can also cause changes in the orbital inclination and/or transit duration.}

%%%%%%%%%%%%%%%%%%%%%%%%%%%%%%%%%%%%%%%%%%%%%%%%%%%%%%%%%%%%%%%%%%%%%%%%%%%%%%%%%%%%%%%%%%%%%%%%%%%%%%%%%%%%%%%%%%%%%%%%%%%%%%%%%%%%%%%%%%%%%%%%%%%%%

\section{Observations}
\label{sec:obs}

\subsection{Danish Telescope}

\begin{table*} \centering
\caption{\label{tab:obslog} Log of the Danish Telescope observations obtained for HATS-18. $N_{\rm obs}$ is the number of observations, $T_{\rm exp}$ is the
exposure time, $T_{\rm dead}$ is the dead time between exposures, `Moon illum.' is the fractional illumination of the Moon at the midpoint of the transit, and
$N_{\rm poly}$ is the order of the polynomial fitted to the out-of-transit data. The aperture radii are target aperture, inner sky and outer sky, respectively.}
\setlength{\tabcolsep}{4pt}
\begin{tabular}{lcccccccccccc} \hline
Date of first & Start time & End time  &$N_{\rm obs}$ & $T_{\rm exp}$ & $T_{\rm dead}$ & Filter & Airmass &  Moon        & Aperture radii   & $N_{\rm poly}$ & Scatter \\
observation   &    (UT)    &   (UT)    &              & (s)           & (s)            &        &         & illumination & (pixels) &                & (mmag)  \\
\hline
2017/05/13 & 23:07 & 01:03 &  48 & 100--150 & 13 & $R$ & 1.08 $\to$ 1.00            & 0.913 & 16 24 45 & 1 & 1.05 \\ % Rabus, Evans                     partial(weather)
2017/05/29 & 00:34 & 04:14 & 117 & 100      & 13 & $R$ & 1.01 $\to$ 1.83            & 0.156 & 13 21 45 & 1 & 1.03 \\ % Peixinho,Campbell-White,Sajadian good
2017/06/09 & 23:29 & 02:36 & 100 & 100      & 13 & $R$ & 1.00 $\to$ 1.42            & 0.996 & 11 18 40 & 1 & 1.56 \\ % Campbell-White, Figuera Jaimes   partial (tech)
2017/06/13 & 23:26 & 01:58 &  81 & 80--100  & 13 & $R$ & 1.01 $\to$ 1.33            & 0.825 & 12 20 50 & 1 & 1.78 \\ % Rahvar, Figuera Jaimes           iffy (cloud)
2017/06/18 & 23:31 & 03:12 & 110 & 100      & 13 & $R$ & 1.03 $\to$ 2.07            & 0.331 & 13 20 40 & 1 & 1.15 \\ % Rahvar, von Essen                good
2017/07/09 & 22:49 & 02:00 &  91 & 100      & 25 & $R$ & 1.09 $\to$ 2.22            & 0.993 & 11 18 40 & 2 & 1.19 \\ % Snodgrass, Rommel, Martino       good
2018/04/24 & 23:34 & 03:31 & 123 & 100      & 14 & $R$ & 1.20 $\to$ 1.00 $\to$ 1.06 & 0.735 & 10 20 40 & 2 & 1.06 \\ % Rabus                            good
2018/04/29 & 23:34 & 04:13 & 140 & 100      & 14 & $R$ & 1.14 $\to$ 1.00 $\to$ 1.18 & 0.998 &  9 15 30 & 2 & 1.33 \\ % Rabus, Hinse                     good
2018/05/05 & 00:50 & 04:32 & 117 & 100      & 14 & $R$ & 1.01 $\to$ 1.00 $\to$ 1.31 & 0.768 & 10 20 40 & 1 & 1.15 \\ % Peixinho, Hinse                  good
2018/05/20 & 23:08 & 02:58 & 118 & 100      & 16 & $R$ & 1.04 $\to$ 1.00 $\to$ 1.20 & 0.374 & 10 18 40 & 1 & 1.28 \\ % Longa-Peña                       good
2018/05/25 & 23:19 & 03:30 & 125 & 100      & 16 & $R$ & 1.01 $\to$ 1.00 $\to$ 1.41 & 0.874 & 10 18 50 & 1 & 2.45 \\ % Longa-Peña, Rahvar               OK (hazy sky)
2018/06/20 & 22:50 & 02:30 & 109 & 100      & 14 & $R$ & 1.01 $\to$ 1.69            & 0.564 & 14 22 50 & 1 & 1.36 \\ % Campbell-White, Rahvar           OK (clouds)
2019/05/17 & 01:39 & 05:14 & 108 & 100      & 16 & $R$ & 1.03 $\to$ 1.94            & 0.961 &  9 18 40 & 1 & 1.37 \\ % Peixinho                         good
2019/05/22 & 22:56 & 01:40 &  82 & 85--100  & 15 & $R$ & 1.03 $\to$ 1.00 $\to$ 1.02 & 0.826 &  9 16 35 & 1 & 1.20 \\ % Tregloan-Reed, Rabus             OK (techissues)
2019/06/01 & 23:09 & 02:43 & 133 & 60--100  & 16 & $R$ & 1.01 $\to$ 1.00 $\to$ 1.31 & 0.023 &  9 16 40 & 2 & 1.22 \\ % Longa-Peña, Salas                good
2021/05/14 & 23:19 & 03:46 & 103 & 100      & 28 & $R$ & 1.02 $\to$ 1.00 $\to$ 1.04 & 0.094 &  8 15 30 & 2 & 1.13 \\ % Korhonen                         good
2021/05/25 & 01:43 & 02:55 & 111 & 100      & 15 & $R$ & 1.02 $\to$ 2.25            & 0.972 &  9 16 35 & 2 & 1.52 \\ % Tregloan-Reed                    good
2021/06/04 & 22:48 & 02:15 & 106 & 100      & 18 & $R$ & 1.01 $\to$ 1.00 $\to$ 1.26 & 0.246 & 13 20 40 & 2 & 1.80 \\ % Sajadian, Khalouei               good
2021/06/09 & 23:37 & 02:53 &  99 & 100      & 18 & $R$ & 1.01 $\to$ 1.00 $\to$ 1.54 & 0.002 & 10 16 35 & 2 & 1.43 \\ % Sajadian, Figuera Jaimes         good
2021/06/14 & 23:32 & 03:30 &  80 & 100      & 18 & $R$ & 1.02 $\to$ 1.00 $\to$ 2.09 & 0.192 & 14 20 40 & 1 & 2.34 \\ % Khalouei, Figuera Jaimes         clouds
\hline \end{tabular} \end{table*}

\begin{figure*} \includegraphics[width=\textwidth,angle=0]{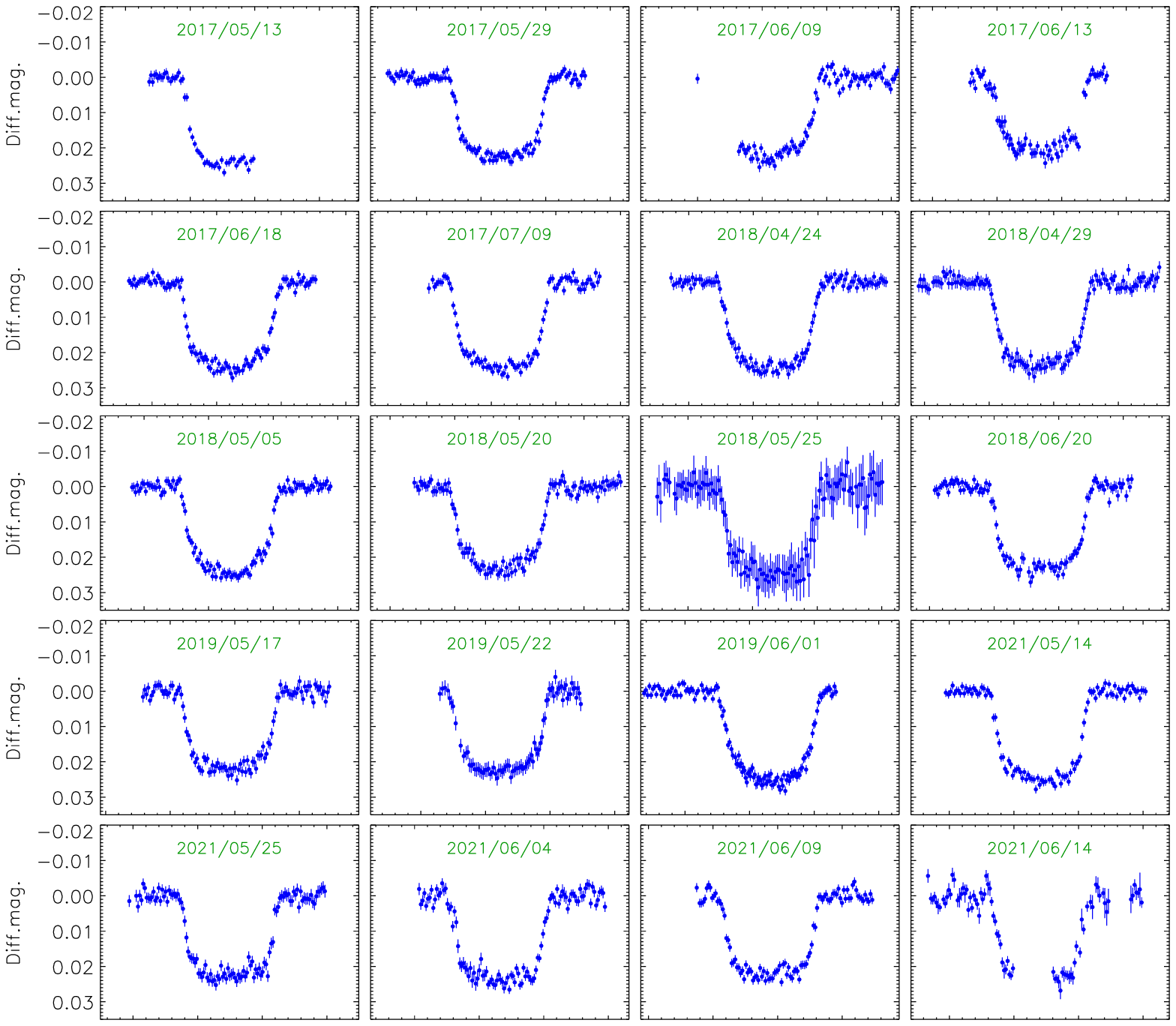}
\caption{\label{fig:dk} Light curves of HATS-18 obtained using the Danish
telescope. In each case the $x$-axis shows 0.2\,d centred on the time of minimum
measured for that transit, but the BJDs are not shown for clarity. The date of
each observation is labelled on the plot.} \end{figure*}

HATS-18 was observed in the 2017, 2018, 2019 and 2021 observing seasons using the Danish 1.54\,m telescope and DFOSC imager at ESO La Silla. No observations were performed in 2020 due to the Covid-19 pandemic. In all cases a Cousins $R$ filter was used to maximise the photon rate for this relatively faint star ($V=14.1$), the telescope was operated out of focus to improve the photometric precision \citep{Me+09mn,Me+09mn2}. \reff{DFOSC has a field of view of 13.7\am$\times$13.7\am\ at a plate scale of 0.39\as\ per pixel, but the CCD was typically windowed down to 12\am$\times$8\am\ to decrease the readout time.} The autoguider was not used because of technical issues.

A total of 20 transits were observed and an observing log is given in Table\,\ref{tab:obslog}. Of the 20 transits, 13 were good, five were affected by atmospheric cloud or haze (2017/05/13, 2017/06/13, 2018/05/25, 2018/06/20, 2021/06/14) and two suffered from technical problems (2017/06/09, 2019/05/22).

Data reduction was performed using the {\sc defot} pipeline \citep{Me+09mn,Me+14mn}, which in turn uses routines from the NASA {\sc astrolib} library\footnote{{\tt http://idlastro.gsfc.nasa.gov/}} {\sc idl}\footnote{{\tt http://www.harrisgeospatial.com/SoftwareTechnology/ IDL.aspx}} implementation of the {\sc aper} routine from {\sc daophot} \citep{Stetson87pasp}. Image motion was \reff{measured} by cross-correlating each image with a reference image obtained \reff{near} the midpoint of \reff{that} observing sequence. The software apertures were placed by hand on the reference images and the radii of the aperture and sky annulus were chosen manually. No bias or flat-field calibrations were applied, as we found that these had little effect beyond increasing the scatter of the results. Cloud-affected data were rejected when they gave measurements that were unreliable.

Differential-magnitude light curves were constructed by optimising the weights of between two and five comparison stars simultaneously with a low-order polynomial fitted to the out-of-transit data (Table\,\ref{tab:obslog}). We consistently found that the choices taken during data reduction (bias, flat-field, aperture size, comparison stars, reference image) had little effect on the shape of the transit in the light curve but \reff{could} have a significant effect on the scatter. The data are plotted in Fig.\,\ref{fig:dk}. We have placed them on the BJD(TDB) timescale using routines from \citet{Eastman++10pasp}.

\subsection{Jongen Telescope}

Nine transits of HATS-18 were observed in the 2020 season using the Yves Jongen Telescope at Deep Sky Chile. This is a PlaneWave Corrected Dall-Kirkham telescope with an aperture of 430\,mm, mounted on a L500 Plane Wave Mount, and equipped with a Moravian 4G CCD camera and EFW 4L-7 filter wheel. \reff{This setup has a field of view of 21.0\am$\times$15.8\am\ at a plate scale of 0.63\as\ per pixel. A Chroma imaging IR-UV filter was used, which has a high transmission between 370 and 700\,nm.}

The data reduction was performed using the Munipack\footnote{\texttt{https://munipack.physics.muni.cz/}} software. This comprised standard calibrations and brightness measurement using aperture photometry. \reff{Differential-magnitude light curves were calculated versus five nearby comparison stars.}

\subsection{Transiting Exoplanet Survey Satellite}

% Sector 10 (2019-Mar-26 to 2019-Apr-22, in cycle 1): observed in camera 1.
% Sector 36 (2021-Mar-07 to 2021-Apr-02, in cycle 3): observed in camera 1.
% Sector 63 (2023-Mar-10 to 2023-Apr-06, in cycle 5): observed in camera 1.

% \begin{figure} \includegraphics[width=\columnwidth,angle=0]{plotTESS.eps}
% \caption{\label{fig:tess} Light curves of HATS-18 from TESS. The data have been
% converted into orbital phase and binned into bins of length 0.002 phase units
% for display purposes. The two sectors (labelled) are shown separately.} \end{figure}

\begin{figure*} \includegraphics[width=\textwidth,angle=0]{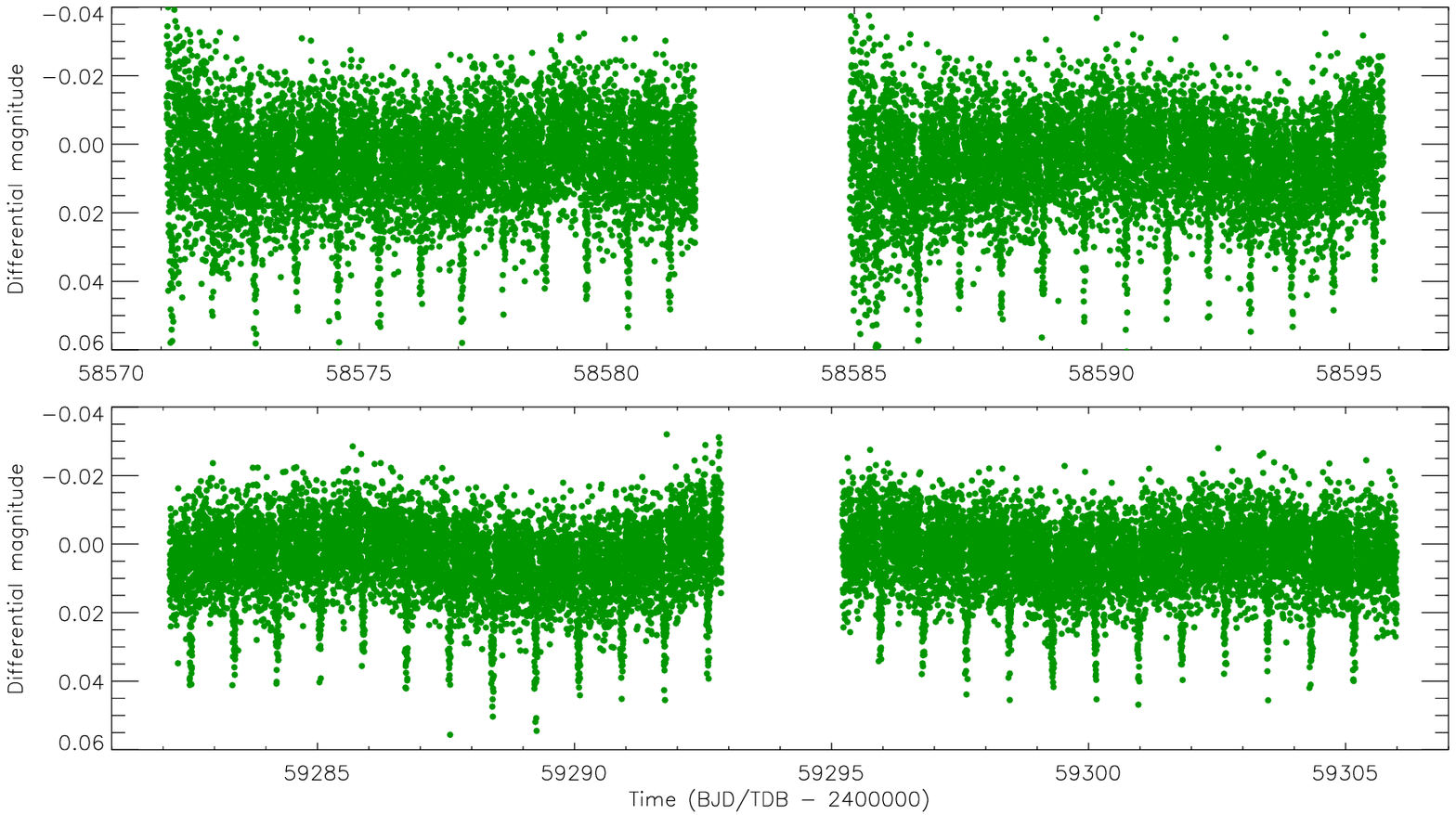}
\caption{\label{fig:tess} PDCSAP light curves of HATS-18 from TESS sectors
10 (top) and 36 (bottom).} \end{figure*}

TESS \citep{Ricker+15jatis} observed HATS-18 in sectors 10 (2019/03/26 to 2019/04/22) and 36 (2021/03/07 to 2021/04/02). In both sectors it was selected as a short-cadence target. One further observation of HATS-18 with TESS is currently scheduled, in sector 69 (2023 March). We downloaded the available observations from the Mikulski Archive for Space Telescopes (MAST) archive\footnote{\texttt{https://mast.stsci.edu/portal/Mashup/Clients/Mast/ Portal.html}} and converted them into magnitude units.

We decided to use the pre-search data conditioning (PDCSAP) light curves \citep{Jenkins+16spie} rather than the simple aperture photometry (SAP) observations, as they have a slightly lower scatter. We retained only those datapoints with a QUALITY flag of zero: 15\,390 in sector 10 and 15\,449 in sector 36. We further rejected all datapoints more than 0.24\,d (three times the transit duration) from the mindpoint of a transit (approximately 71\% of the data) as these were not useful for our analysis. The data are quite scattered due to the relative faintness of HATS-18, and the small aperture and large pixel scale of TESS (Fig.\,\ref{fig:tess}).

%%%%%%%%%%%%%%%%%%%%%%%%%%%%%%%%%%%%%%%%%%%%%%%%%%%%%%%%%%%%%%%%%%%%%%%%%%%%%%%%%%%%%%%%%%%%%%%%%%%%%%%%%%%%%%%%%%%%%%%%%%%%%%%%%%%%%%%%%%%%%%%%%%%%%%%%%%%%%%%%%%%%%%%%%%%%%%%%%%%%%%%%%%%%%%%%%%%%%%%

\section{Physical properties of HATS-18}
\label{sec:absdim}

\begin{table*} \centering \caption{\label{tab:absdim} Derived physical properties
of the HATS-18 system. When measurements are accompanied by two errorbars, the
first refers to the statistical uncertainty and the second to the systematic
uncertainty. The properties from \citet{Penev+16aj} are included for comparison.}
\begin{tabular}{lllcc}
\hline
Parameter                     & Symbol                & Unit      & This work                         & \citet{Penev+16aj}                 \\
\hline
Orbital period                & $P$                   & d         & $0.83784369 \pm 0.00000011$       & $0.83784340 \pm 0.00000047$        \\
Sum of the fractional radii   & $r_{\rm A}+r_{\rm b}$ &           & $0.2979 \pm 0.0043$               &                                    \\
Ratio of the radii            & $k$                   &           & $0.13426 \pm 0.00095$             & $0.1347 \pm 0.0019$                \\
Orbital inclination           & $i$                   & $^\circ$  & $88.2 \pm 1.8$                    & $85.5^{+1.9}_{-2.8}$               \\
Fractional radius of star     & $r_{\rm A}$           &           & $0.2627 \pm 0.0036$               & $0.2695^{+0.0090}_{-0.0078}$       \\
Fractional radius of planet   & $r_{\rm b}$           &           & $0.03524 \pm 0.00065$             &                                    \\
\hline
Stellar effective temperature & \Teff                 & K         &                                   & $5600 \pm 120$                     \\
Stellar metallicity           & [Fe/H]                & dex       &                                   & $+0.280 \pm 0.080$                 \\
Stellar velocity amplitude    & $K_{\rm A}$           & \ms       &                                   & $415.2 \pm 10.0$                   \\
Stellar mass                  & $M_{\rm A}$           & \Msun     & $1.048 \pm 0.057 \pm 0.027$       & $1.037 \pm 0.047$                  \\
Stellar radius                & $R_{\rm A}$           & \Rsun     & $0.999 \pm 0.021 \pm 0.009$       & ${1.020}^{+0.057}_{-0.031}$        \\
Stellar surface gravity       & $\log g_{\rm A}$      & (cgs)     & $4.460 \pm 0.015 \pm 0.004$       & $4.436 \pm 0.034$                  \\
Stellar density               & $\rho_{\rm A}$        & \psun     & $1.052 \pm 0.043$                 & ${0.978}^{+0.092}_{-0.149}$        \\[2pt]
Planetary mass                & $M_{\rm b}$           & \Mjup     & $1.990 \pm 0.087 \pm 0.034$       & $1.980 \pm 0.077$                  \\
Planetary radius              & $R_{\rm b}$           & \Rjup     & $1.304 \pm 0.034 \pm 0.011$       & ${1.337}^{+0.102}_{-0.049}$        \\
Planetary surface gravity     & $g_{\rm b}$           & \mss      & $29.0 \pm 1.3$                    & ${27.2}^{+2.3}_{-3.7}$             \\
Planetary density             & $\rho_{\rm b}$        & \pjup     & $0.840 \pm 0.053 \pm 0.007$       & ${0.769}^{+0.098}_{-0.151}$        \\[2pt]
Equilibrium temperature       & \Teq                  & K         & $2029 \pm 45$                     & $2060 \pm 59$                      \\
Safronov number               & \safronov\            &           & $0.0514 \pm 0.0018 \pm 0.0004$    & ${0.0498}^{+0.0025}_{-0.0033}$     \\
Semimajor axis                & $a$                   & au        & $0.01768 \pm 0.00032 \pm 0.00015$ & $0.01761 \pm 0.00027$              \\
Age                           & $\tau$                & Gyr       & $3.2^{+2.2\,+1.7}_{-2.1\,-1.3}$   & $4.2 \pm 2.2$                      \\
\hline \end{tabular} \end{table*}

The only existing measurement of the physical properties of the HATS-18 system is by \citet{Penev+16aj}, who had access to only one high-quality transit light curve of the system. We have therefore redetermined the physical properties of HATS-18 using our extensive photometry and the methods from the \textit{Homogeneous Studies} project \citep[][and references therein]{Me12mn}. The measured parameters are useful in subsequent analyses.

All modelling of the light curves of HATS-18 was done using version 42 of the {\sc jktebop}\footnote{{\sc jktebop} is written in {\sc fortran77} and the source code is available at {\tt http://www.astro.keele.ac.uk/jkt/codes/jktebop.html}} code \citep[][and references therein]{Me13aa}. We first fitted each transit from the Danish Telescope individually, to check the level of scatter in the light curves. The fitted parameters were the transit midpoint ($T_0$), the fractional radii of the two components ($r_{\rm A}=\frac{R_{\rm A}}{a}$ and $r_{\rm b}=\frac{R_{\rm b}}{a}$ where $R_{\rm A}$ is the radius of the star, $R_{\rm b}$ is the radius of the planet, and $a$ is the semimajor axis of the relative orbit) expressed as their sum and ratio, and the orbital inclination ($i$).

Once this was done, we merged the data from the 17 complete transits into a single light curve file, at the same time scaling the point errors in each dataset to obtain $\chir=1$ versus the best fit. We did not use the three partially-observed transits as they are less reliable than the others. We also did not use the transits observed with the Jongen telescope or TESS due to their higher scatter. We obtained a best model to the data by fitting for $T_0$, $P$, $r_{\rm A}+r_{\rm b}$, $k$, $i$ and a polynomial function for each transit (see Table\,\ref{tab:obslog}). This was done for four different two-parameter limb darkening (LD) laws \citep[see][]{Me08mn} and with either both coefficients fixed or the linear coefficient fitted. Fixed and initial values of the LD coefficients were taken from the theoretical tabulations by \citet{Claret00aa,Claret04aa2,Claret17aa}.

The uncertainties in the fitted parameters were calculated using Monte Carlo and residual-permutation simulations \citep[see][]{Me08mn}. The adopted value of each fitted parameter is the weighted mean of the four fits with one LD coefficient fitted. Its uncertainty is the larger of the Monte Carlo and residual-permutation options, with an additional contribution from the variation in parameter values added in quadrature. These results are given in the top portion of Table\,\ref{tab:absdim}.

To determine the physical properties of the system an additional constraint is needed \citep{Me09mn} for which we resorted to interpolating within tabulated predictions of stellar properties from theoretical models. Measurements of the spectroscopic properties of the host star (temperature \Teff, metallicity [Fe/H], velocity amplitude $K_{\rm A}$) were taken from \citet{Penev+16aj}. We estimated an initial value of the velocity amplitude of the planet $K_{\rm b}$ and used the measured values of $K_{\rm A}$, $i$, $P$, $r_{\rm A}$ and $r_{\rm b}$ to determine the physical properties of the system. This process was iterated to find the value of $K_{\rm b}$ that gave the best match between the observed and predicted \Teff\ and $r_{\rm A}$. We included [Fe/H] as a constraint and performed a grid search over age to determine the overall best set of system properties. This process was undertaken for five different sets of theoretical stellar evolutionary models \citep{Claret04aa,Demarque+04apjs,Pietrinferni+04apj,Vandenberg++06apjs,Dotter+08apjs}.

Random errors were propagated using a perturbation approach and systematic errors were estimated from the differences between the five sets of solutions using the various theoretical predictions. Our final set of properties of the system are collected in Table\,\ref{tab:absdim}, which also shows the values from \citet{Penev+16aj} for comparison. Our new observations have enabled, amongst other results, significantly better measurements of $r_{\rm A}$ (a parameter important for tidal evolution) and $\rho_{\rm b}$ (relevant in understanding the structure and evolution of planets). The age determination remains frustratingly uncertain; an improved value might be obtained from a more precise \Teff\ measurement. Although HATS-18\,A has a measured rotation period of $9.8 \pm 0.4$\,d \citep{Penev+16aj} indicative of a relatively young age, there is evidence that gyrochronological ages are unreliable for the host stars of hot Jupiters \citep{PoppenhaegerWolk14aa,Brown14mn,Maxted++15aa2,Mancini+17mn}.

%%%%%%%%%%%%%%%%%%%%%%%%%%%%%%%%%%%%%%%%%%%%%%%%%%%%%%%%%%%%%%%%%%%%%%%%%%%%%%%%%%%%%%%%%%%%%%%%%%%%%%%%%%%%%%%%%%%%%%%%%%%%%%%%%%%%%%%%%%%%%%%%%%%%%

\section{Transit timing analysis}

\subsection{Measurement of the transit times}
\label{sec:tmin}

\begin{table*} \centering \caption{\label{tab:tmin}
All times of mid-transit for HATS-18 used in the current work.}
\begin{tabular}{r@{\,$\pm$\,}lrrl} \hline
\mc{BJD(TDB)} & Cycle & Residual (s) & Source \\             % Observer(s)
\hline
2457410.80009 & 0.00030 & $-$1451.0 &    0.00024 & This work (\citealt{Penev+16aj} transit) \\   %    20.9 s    0.6 sigma     Penev+2016_my_measurement_TDB
2457834.74845 & 0.00040 &  $-$945.0 & $-$0.00037 & \citet{Patra+20aj}                       \\   %   -31.9 s   -0.7 sigma     Patra+2020
2457902.61444 & 0.00019 &  $-$864.0 &    0.00027 & This work (Danish telescope)             \\   %    23.5 s    1.1 sigma     Danish-20170528
2457918.53333 & 0.00063 &  $-$845.0 &    0.00013 & This work (Danish telescope)             \\   %    11.2 s    0.2 sigma     Danish-20170613
2457923.55998 & 0.00017 &  $-$839.0 & $-$0.00028 & This work (Danish telescope)             \\   %   -24.5 s   -1.3 sigma     Danish-20170619
2457944.50658 & 0.00018 &  $-$814.0 &    0.00022 & This work (Danish telescope)             \\   %    19.1 s    0.9 sigma     Danish-20170709
2458217.64370 & 0.00029 &  $-$488.0 &    0.00026 & \citet{Patra+20aj}                       \\   %    22.2 s    0.7 sigma     Patra+2020
2458233.56286 & 0.00021 &  $-$469.0 &    0.00038 & This work (Danish telescope)             \\   %    33.2 s    1.4 sigma     Danish-20180424
2458238.58946 & 0.00027 &  $-$463.0 & $-$0.00008 & This work (Danish telescope)             \\   %    -6.8 s   -0.2 sigma     Danish-20180429
2458243.61625 & 0.00018 &  $-$457.0 & $-$0.00035 & This work (Danish telescope)             \\   %   -30.4 s   -1.5 sigma     Danish-20180504
2458259.53486 & 0.00027 &  $-$438.0 & $-$0.00077 & This work (Danish telescope)             \\   %   -66.9 s   -2.2 sigma     Danish-20180520
2458264.56344 & 0.00068 &  $-$432.0 &    0.00074 & This work (Danish telescope)             \\   %    64.2 s    0.8 sigma     Danish-20180525
2458290.53560 & 0.00024 &  $-$401.0 & $-$0.00025 & This work (Danish telescope)             \\   %   -22.0 s   -0.8 sigma     Danish-20180620
2458581.26804 & 0.00043 &   $-$54.0 &    0.00038 & This work (TESS)                         \\   %    32.9 s    0.7 sigma     TESS_sector10
2458620.64595 & 0.00021 &    $-$7.0 & $-$0.00037 & This work (Danish telescope)             \\   %   -31.8 s   -1.3 sigma     Danish-20190516
2458626.51102 & 0.00045 &       0.0 & $-$0.00021 & This work (Danish telescope)             \\   %   -17.7 s   -0.3 sigma     Danish-20190522
2458636.56579 & 0.00020 &      12.0 &    0.00044 & This work (Danish telescope)             \\   %    37.9 s    1.7 sigma     Danish-20190601
2458898.81064 & 0.00036 &     325.0 &    0.00017 & This work (Jongen telescope)             \\   %    15.1 s    0.4 sigma     YvesJongen
2458862.78435 & 0.00068 &     282.0 &    0.00117 & This work (Jongen telescope)             \\   %   101.0 s    1.3 sigma     YvesJongen
2458883.72975 & 0.00048 &     307.0 &    0.00047 & This work (Jongen telescope)             \\   %    40.9 s    0.8 sigma     YvesJongen
2458903.83713 & 0.00054 &     331.0 & $-$0.00040 & This work (Jongen telescope)             \\   %   -34.4 s   -0.6 sigma     YvesJongen
2458950.75579 & 0.00088 &     387.0 & $-$0.00099 & This work (Jongen telescope)             \\   %   -85.7 s   -0.9 sigma     YvesJongen
2458972.54162 & 0.00046 &     413.0 &    0.00090 & This work (Jongen telescope)             \\   %    77.7 s    1.5 sigma     YvesJongen
2458982.59355 & 0.00078 &     425.0 & $-$0.00130 & This work (Jongen telescope)             \\   %  -112.0 s   -1.3 sigma     YvesJongen
2459034.54221 & 0.00069 &     487.0 &    0.00105 & This work (Jongen telescope)             \\   %    90.4 s    1.2 sigma     YvesJongen
2459292.59679 & 0.00029 &     795.0 & $-$0.00027 & This work (TESS)                         \\   %   -23.2 s   -0.7 sigma     TESS_sector36
2459203.78668 & 0.00079 &     689.0 &    0.00107 & This work (Jongen telescope)             \\   %    92.1 s    1.0 sigma     YvesJongen
2459349.57034 & 0.00020 &     863.0 & $-$0.00010 & This work (Danish telescope)             \\   %    -8.5 s   -0.4 sigma     Danish-20210514
2459359.62471 & 0.00021 &     875.0 &    0.00015 & This work (Danish telescope)             \\   %    12.6 s    0.5 sigma     Danish-20210524
2459370.51596 & 0.00040 &     888.0 & $-$0.00057 & This work (Danish telescope)             \\   %   -49.5 s   -1.1 sigma     Danish-20210604
2459375.54404 & 0.00042 &     894.0 &    0.00044 & This work (Danish telescope)             \\   %    38.3 s    0.8 sigma     Danish-20210609
2459380.57024 & 0.00058 &     900.0 & $-$0.00042 & This work (Danish telescope)             \\   %   -36.2 s   -0.6 sigma     Danish-20210614
\hline \end{tabular} \end{table*}

We modelled the observed transit light curves individually as detailed in Section\,\ref{sec:absdim}, fitting for $T_0$, $r_{\rm A}+r_{\rm b}$, $k = \frac{r_{\rm b}}{r_{\rm A}}$, $i$ and the linear LD coefficient of the quadratic LD law. We also included a quadratic polynomial to model the out-of-eclipse brightness of the system, in order to capture the uncertainties in the light curve normalisation process at the data-reduction stage. \reff{The two transits from the Danish telescope lacking coverage of either the ingress or egress} (2017/05/13 and 2017/06/09) were not included because such data do not give reliable transit times \citep[e.g.][]{Gibson+09apj}. Uncertainties on the transit times were obtained using Monte Carlo and residual-permutation simulations \citep{Me08mn} and the larger of the two options was used. In cases where the reduced $\chi^2$ of the fit was $\chir > 1$ the errorbar was further multiplied by $\sqrt{\chir}$ to avoid underestimation of the uncertainties.

The TESS data demanded a different approach as the scatter is high. We therefore fitted all transits from each sector simultaneously to determine a single effective time of transit for that sector. The reference time of transit was chosen to be the observed transit closest to the midpoint of the data in each case, and a quadratic polynomial was included to account for slow drifts in the instrumental magnitudes of the system over the sector. The uncertainties were obtained using Monte Carlo simulations.

\subsection{Published transit times}

\citet[][their table 4]{Penev+16aj} presented a single transit time, $T_C$, from a joint analysis of all the light and radial velocity curves available to them. We did not wish to use this directly as it is measured from data taken over four years so is not suitable for any analysis where the orbital period may be changing. We therefore modelled the one follow-up light curve which has full coverage of a transit, using the same approach as above. This was taken on the night of 2016/01/22 using an LCOGT\footnote{\texttt{https://lco.global/}} 1-m telescope and Sinistro imager at the Cerro Tololo Inter-American Observatory (CTIO\footnote{\texttt{https://noirlab.edu/public/programs/ctio/}}). The data were reported as BJD on the UTC time system so we converted the resulting transit time to TDB.

We subsequently found that this transit time occurred significantly later than expected (approximately 100\,s) from a preliminary orbital ephemeris based on our Danish Telescope data. Dr.\ Joel Hartman kindly made available the original data for us to reduce using our own {\sc defot} code. We found that our own light curve has timestamps in excellent agreement with those from \citet{Penev+16aj}, even though ours are expressed in TDB and the published data are given in UTC. Furthermore, our own transit time is in much better agreement with the preliminary orbital ephemeris. We conclude that the timestamps from the published data are actually in TDB, not UTC. For subsequent analysis we therefore used our measurement of the transit time, from the light curve presented by \citet{Penev+16aj}, under the assumption that the timestamps are in TDB.

Two further transit times of HATS-18 were given by \citet{Patra+20aj}. These are already in BJD/TDB so were added to our analysis without modification. We are not aware of any further source of transit times for this system.

%%%%%%%%%%%%%%%%%%%%%%%%%%%%%%%%%%%%%%%%%%%%%%%%%%%%%%%%%%%%%%%%%%%%%%%%%%%%%%%%%%%%%%%%%%%%%%%%%%%%%%%%%%%%%%%%%%%%%%%%%%%%%%%%%%%%%%%%%%%%%%%%%%%%%

% With the Penev tmin as TDB not UTC:
% Poly order  Minima       Chisqred      T_0 and T_0 error         Period and period error     Quadratic term and error     Cubic term and error        AIC       BIC    rms(s)
% Linear      All minima   1.79047   58626.511225 +/- 0.000071   0.837843816 +/- 0.000000105                                                           61.30     64.23    50.15
% Quadratic   All minima   1.78443   58626.511201 +/- 0.000102   0.837843823 +/- 0.000000107   5.5222e-11 +/- 1.6785e-10                               63.10     67.50    50.25
% Cubic       All minima   1.71471   58626.511285 +/- 0.000125   0.837844044 +/- 0.000000221  -1.2191e-10 +/- 2.2625e-10  -2.898e-13 +/- 2.540e-13     62.87     68.73    49.45

\subsection{Orbital ephemerides}
\label{sec:ttv}

\begin{figure*}
\includegraphics[width=\textwidth,angle=0]{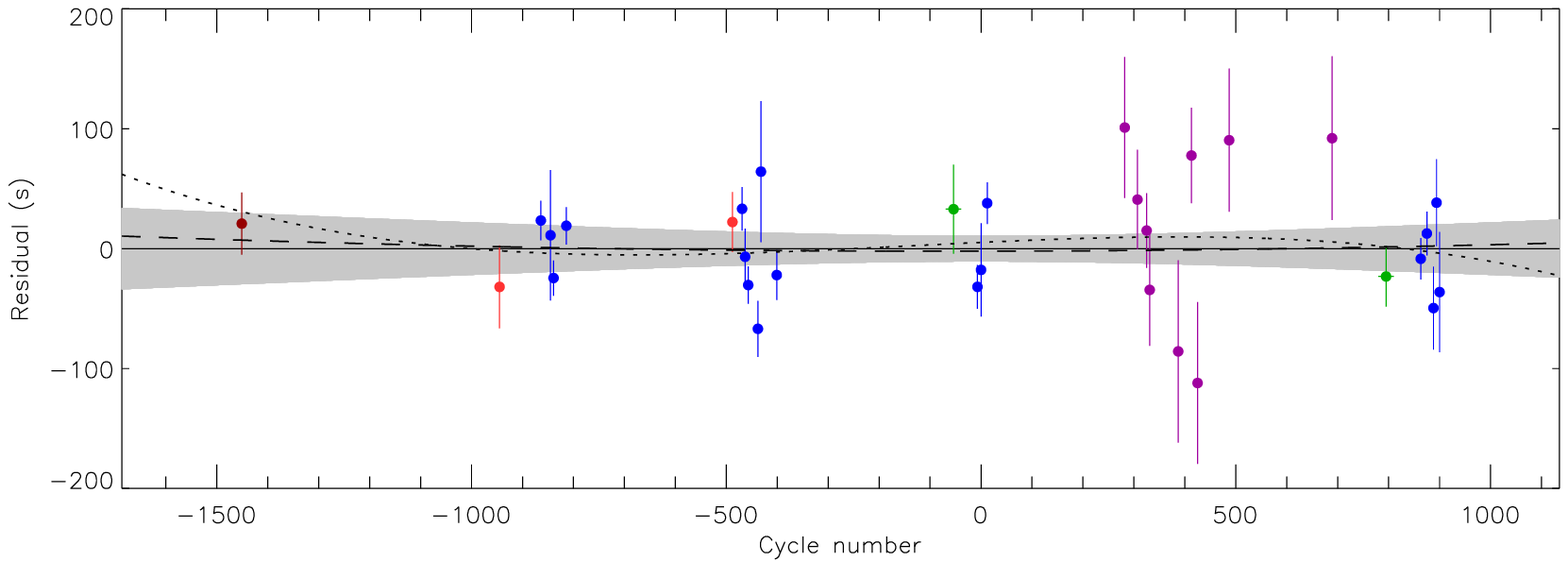}
\caption{\label{fig:minima} \reff{Plot of the residuals of the timings of mid-transit versus a
linear ephemeris. Light red points indicate published timings, the dark red point is our analysis
of the transit light curve from \citet{Penev+16aj}, blue points are timings from the Danish
telescope, purple from the Jongen telescope and green from TESS. The dashed line indicates
the difference between the best-fitting linear and quadratic ephemerides, and the dotted
line the difference between the linear and cubic ephemerides. The grey shade indicates
the uncertainty in the linear ephemeris as a function of orbital cycle.}}
\end{figure*}

\begin{table*} \centering \caption{\label{tab:ephem}
Properties of the ephemerides calculated in this work. The linear ephemeris was adopted as the final one.}
\begin{tabular}{lccc} \hline
Quantity                      & Linear ephemeris      & Quadratic                         & Cubic                             \\
                              & (adopted)             & ephemeris                         & ephemeris                         \\
\hline
Reference time (BJD/TDB)      & 2458626.51123 (7)     & 2458626.51120 (10)                & 2458626.51129 (13)                \\
Linear term (d)               & 0.83784382 (11)       & 0.83784382 (11)                   & 0.83784404 (22)                   \\
Quadratic term ($p_1$)        &                       & (0.6$\pm$1.7) $\times10^{-10}$    & ($-$1.2$\pm$2.3) $\times10^{-10}$ \\
Cubic term ($p_2$)            &                       &                                   & ($-$2.9$\pm$2.5) $\times10^{-13}$ \\
r.m.s.\ of residuals (s)      & 50.2                  & 50.2                              & 49.5                              \\
BIC                           & 64.2                  & 67.5                              & 68.7                              \\
AIC                           & 61.3                  & 63.1                              & 62.9                              \\
\hline \end{tabular} \end{table*}

Once the transit times were assembled we fitted them with a straight line to obtain a linear ephemeris:
$$ T_0 = {\rm BJD(TDB)} \,\, 2458626.511225 (71) \, + \, 0.83784382 (11) \times E $$
where $E$ is the epoch of the transit and the bracketed quantities indicate the uncertainties in the preceding digits. The fit has $\chir = 1.79$, which is significantly larger than the 1.0 expected for a good fit but in line with our past experience that \chir\ is usually more than 1.0 in this kind of analysis \citep[e.g.][]{Me+16mn,Basturk+22mn}. The $1\sigma$ uncertainties in the ephemeris above have been multiplied by $\sqrt{\chir}$ to account for this relatively poor fit. The transit times and their residuals are given in Table\,\ref{tab:tmin}.

Once the linear ephemeris was established we tried fitting for quadratic and cubic ephemerides of the \reff{forms}
$$ T_0 = T_{\rm ref} \, + \, P\,E \, + \, p_1E^2 $$
and
$$ T_0 = T_{\rm ref} \, + \, P\,E \, + \, p_1E^2 + p_2E^3 $$
where $p_1$ and $p_2$ are the coefficients of the quadratic and cubic terms, respectively. These give a very similar fit, with slightly larger values for the Bayesian Information Criterion \citep[BIC;][]{Schwarz78,Liddle07mn} and Akaike Information Criterion \citep[AIC;][]{Akaike81} \reff{than the linear ephemeris}. The properties of these fits are compared with the linear ephemeris in Table\,\ref{tab:ephem} and shown graphically in Fig.\,\ref{fig:minima}. \reff{The uncertainties in the fitted coefficients were determined using 10\,000 Monte Carlo simulations, which were found to be in excellent agreement with the formal errors outputted by the fitting code\footnote{To fit the ephemerides we used the \textsc{poly\_fit} routine in IDL.}.}

The curvature term in the quadratic ephemeris represents a period decrease which could be attributed to tidally-induced orbital decay if it were negative. Instead it can be seen that this term is positive but consistent with zero. The cubic ephemeris is designed to allow the detection of a changing acceleration which could be attributed to orbital motion with a third body in the system. However, the quadratic and cubic ephemerides are not a significantly better fit to the data than the linear ephemeris, so we are only able to set upper limits on orbital period changes. Future monitoring of the system is needed to refine these limits and push the noise down below any real astrophysical signal.

Finally, we used the {\sc period04} package \citep{LenzBreger05coast} to check for \emph{periodic} variations in the residuals of the timings versus the linear ephemeris. No significant signals were found, with the strongest being at 0.0545 cycle$^{-1}$ with a signal to noise ratio of 2.9, well below the commonly-used signal-to-noise ratio threshold of 4.0 \citep{Breger+93aa}. We therefore find no evidence in our data for a periodic variation in the times of transit \reff{for} HATS-18.

\subsection{Constraints on the tidal quality factor}

% Quantity                   Value          errorbar     3-sigma limit     errorbar    unit
% -----------------------   -------------------------    --------------------------    ----------
% Quadratic coefficient =  -4.5000e-10 +/- 0.0000e+00    -4.5000e-10 +/- 0.0000e+00   (day/epoch)
% dP / dE =                -9.0000e-10 +/- 0.0000e+00    -9.0000e-10 +/- 0.0000e+00   (day/epoch)
% dP / dt =                -1.0742e-09 +/- 0.0000e+00    -1.0742e-09 +/- 0.0000e+00   (d/d) or (s/s)
% dP / dt =                -3.3898e-02 +/- 0.0000e+00    -3.3898e-02 +/- 0.0000e+00   (s/yr)
% dP / dt =                -3.3898e+01 +/- 0.0000e+00    -3.3898e+01 +/- 0.0000e+00   (ms/yr)
% Qdash =                   1.2758e+05 +/- 1.2672e+04     1.2758e+05 +/- 1.2430e+04
% log10(Qdash) =            5.1058e+00 +/- 4.2793e-02     5.1058e+00 +/- 4.2160e-02

The tidal quality factor, $Q_\star$, is a measure of the efficiency of the dissipation of tidal energy \citep[e.g.][]{Ogilvie14araa}. To constrain $Q_\star$ we followed the approach from \citet{Birkby+14mn}, which refers to the modified tidal quality factor
$$ Q_\star^{\,\prime} = \frac{3}{2} \, \frac{Q_\star}{k_2} $$
where $k_2$ is the Love number \citep{Love11book}. This is related to measurable properties of the system via the equation
$$ Q_\star^{\,\prime} = \frac{-27}{~~8} \left(\frac{M_{\rm b}}{M_{\rm A}}\right) \left(\frac{R_{\rm A}}{a}\right)^5 \left(\frac{2\pi}{\Porb}\right) \frac{1}{p_1} $$
\citep{Wilkins+17apj} where $(R_{\rm A}/a)$ is simply $r_{\rm A}$, $p_1$ is the quadratic coefficient in the ephemeris, and the other terms have the meanings given in Table\,\ref{tab:absdim}.

Our value of $p_1$ in Table\,\ref{tab:ephem} is greater than zero (i.e.\ it indicates an increasing orbital period) so would cause a negative $Q_\star^{\,\prime}$ if inserted blindly into the equation above. We therefore used its $3\sigma$ lower limit of $p_1^{\rm lim} = -4.5 \times 10^{-10}$ to obtain the constraint that $Q_\star^{\,\prime} > 10^{5.11 \pm 0.04}$, where the errorbar comes from propagating the uncertainties in $M_{\rm A}$, $M_{\rm b}$ and $r_{\rm A}$. This limit is comparable to previous constraints for other systems \citep[e.g.][]{OgilvieLin07apj,Jackson++08apj2,Penev+12apj}.

%%%%%%%%%%%%%%%%%%%%%%%%%%%%%%%%%%%%%%%%%%%%%%%%%%%%%%%%%%%%%%%%%%%%%%%%%%%%%%%%%%%%%%%%%%%%%%%%%%%%%%%%%%%%%%%%%%%%%%%%%%%%%%%%%%%%%%%%%%%%%%%%%%%%%

\subsection{Comparison of the tidal quality factor with models for stellar tidal dissipation}
\label{sec:barker}

HATS-18 is a 1.05\Msun\ G-star and so it likely harbours a radiative core on the main sequence. It also rotates much slower than the planetary orbit: based on $v\sin i = 6.23$\kms\ \citep{Penev+16aj}, we obtain a lower bound on the stellar rotation period of approximately 8\,d \citep[cf.\ $9.8$\,d in][]{Penev+16aj}. Hence, the dominant tidal mechanism is expected to be excitation and dissipation of internal gravity waves launched from the radiative/convective interface into the radiative core \citep[e.g.][]{GoodmanDickson98apj,OgilvieLin07apj,BarkerOgilvie10mn,Chernov++17mn,Barker20mn}. Hydrodynamical simulations show that if these waves have large enough amplitudes to break, they are efficiently damped \citep{BarkerOgilvie10mn,Barker11mn}, so we may then estimate tidal dissipation rates by assuming these waves are completely absorbed. This gives a simple estimate for the stellar tidal quality factor $Q^\prime_\star$ \citep[see e.g.\ eq.\,41 of][]{Barker20mn}, that can be computed using stellar models together with the currently observed planetary mass and orbital period (note: $Q^\prime_\star \propto P_\mathrm{orb}^{8/3}$).

\begin{figure*}
\includegraphics[width=8.5cm,angle=0]{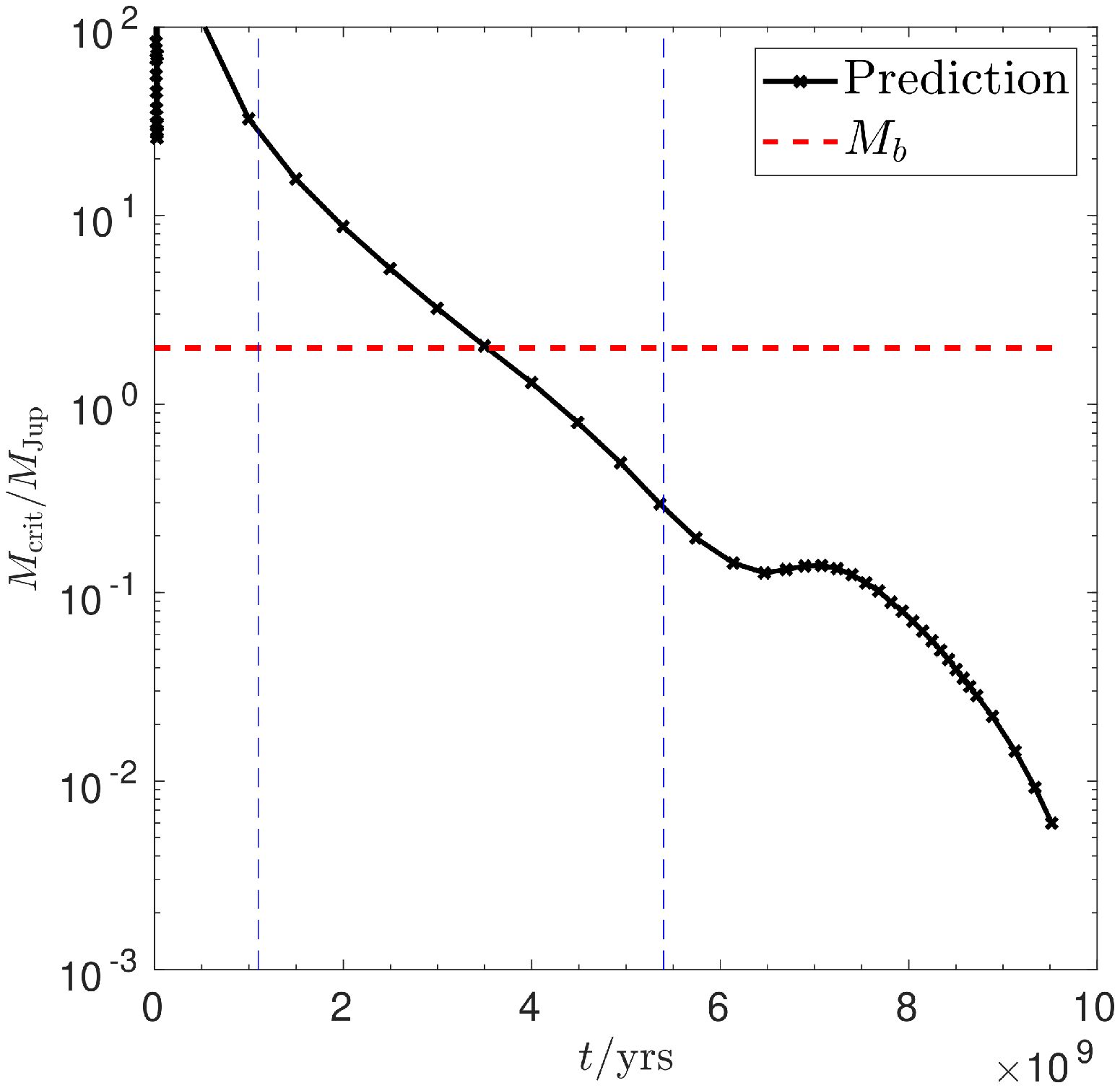} ~~~~
\includegraphics[width=8.5cm,angle=0]{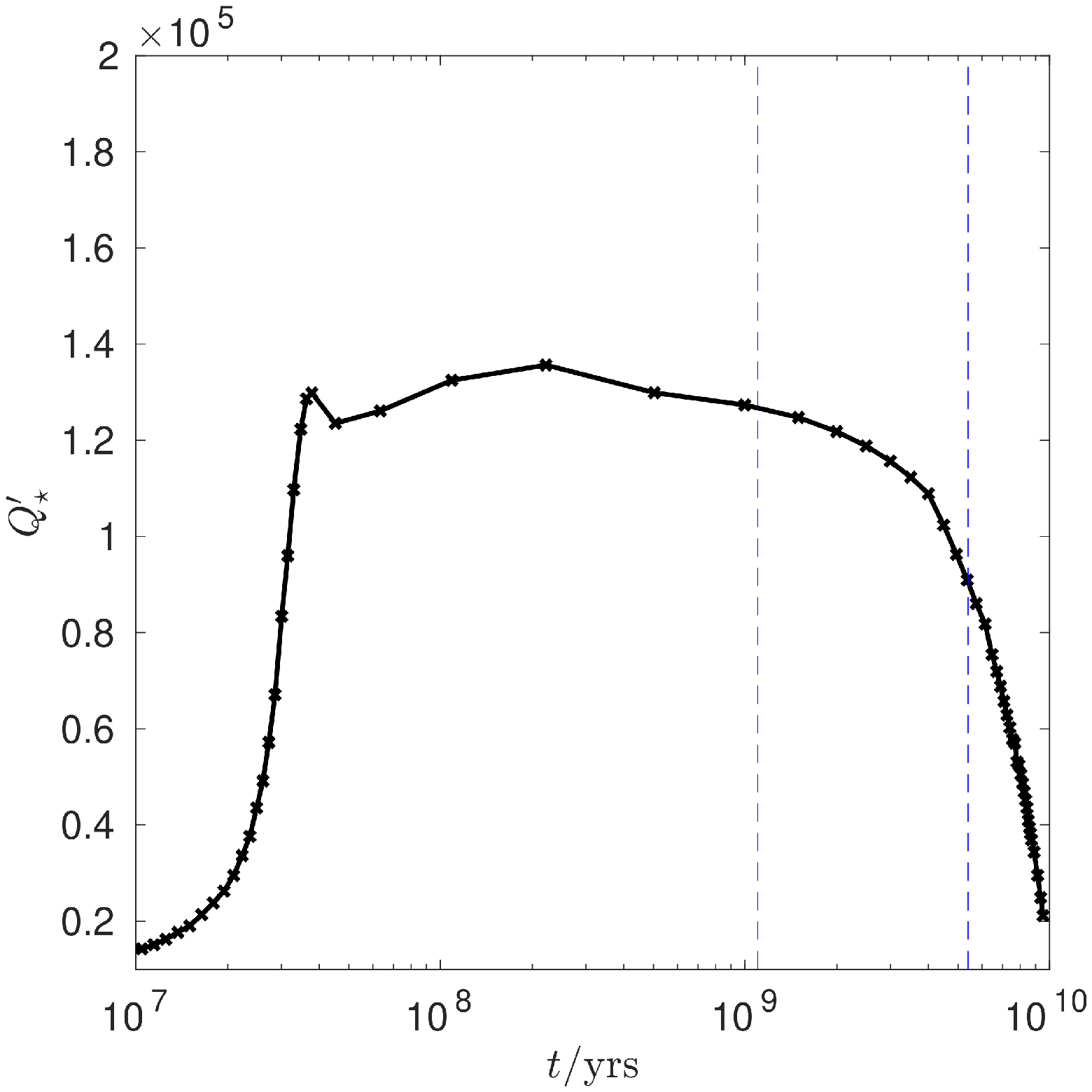} \\
\caption{\label{fig:barker} \textit{Left:} critical planetary mass $M_\mathrm{crit}$ for gravity wave
breaking in \Mjup\ as a function of stellar age $t$ (in yr) in a model of HATS-18. The red dashed line
indicates the mass of HATS-18\,b ($M_{\rm b} = 2$\Mjup). \textit{Right:} $Q^\prime_\star$ due to
tidally-excited gravity waves as a function of stellar age $t$ (in yr), based on the current orbital
period. The blue dashed lines indicate the uncertainty in stellar age in Table\,\ref{tab:absdim}.}
\end{figure*}

The critical planetary mass required to cause wave breaking in the stellar core is predicted to be a strong function of stellar age (and mass). %, but is expected to be satisfied on the main sequence for companions exceeding Jupiter's mass.
To explore this scenario, we computed the evolution of HATS-18 using {\sc mesa} stellar models \citep[r12278, using an initial mass 1.05\Msun\ and metallicity $Z=0.03$;][]{Paxton+11apjs,Paxton+13apjs,Paxton+15apjs,Paxton+18apjs,Paxton+19apjs} and evaluated the critical planetary mass $M_\mathrm{crit}$ required to initiate wave breaking by applying eq.\,47 in \citet{Barker20mn}. Results for $M_\mathrm{crit}/M_{\rm Jup}$ are shown in Fig.\,\ref{fig:barker} (left panel) as a function of stellar age. This shows that wave breaking is not predicted for ages less than 3.5\,Gyr, since $M_\mathrm{crit}$ exceeds the observed planetary mass (red dashed line). For ages older than 3.5\,Gyr, wave breaking is expected in the stellar core, and we show the resulting $Q^\prime_\star$ prediction as a function of age in Fig.\,\ref{fig:barker} (right panel). We typically find $Q^\prime_\star\approx1.2\times 10^5\approx10^{5.08}$ for ages less than approximately 3\,Gyr \citep[similar to the predictions in][]{Barker20mn}. Since we have constrained $Q^\prime_\star \ga 10^{5.11}$, our prediction that $Q’_\star\approx1.5\times 10^5\approx10^{5.08}$ is right at the limit of what can be tested by the available observations.

For earlier ages, or in general for cases where wave breaking is not predicted, gravity waves are likely to be much more weakly damped by radiative diffusion, with correspondingly much larger $Q^\prime_\star$. Alternatively, weaker nonlinear effects such as parametric instabilities could be important, though tidal dissipation rates from these are hard to predict with certainty \citep[e.g.][]{BarkerOgilvie11mn,Weinberg+12apj,EssickWeinberg16apj}. % \textit{The star rotates too slowly for inertial waves to affect the orbital period, and convective damping of equilibrium tides is likely to be very weak \citep[e.g.][]{Duguid+20mn,Barker20mn}, so we expect orbital evolution to be negligible. AJB: omit?}
On the other hand, if it is possible to enter the fully damped regime for gravity waves through processes other than wave breaking (as predicted above), theory would again predict $Q^\prime_\star\approx10^{5.08}$. Such processes include gradual spin-up of the core due to radiative diffusion \citep{BarkerOgilvie10mn}, and perhaps by wave breaking (for lower tidal amplitudes) initiated by passage through a resonance \citep[e.g.][since wave breaking would likely prohibit resonance locking]{MaFuller21apj}. This system therefore remains a very interesting one for follow-up studies with further observations having a strong potential to test tidal theory.

% \textit{Tidal spin-up of the star has also been suggested, which is hard to reconcile with gravity wave damping (other tidal mechanisms are probably negligible) unless the mechanisms of angular momentum transport within the star are particularly efficient. \ \ AJB: omit?}

%%%%%%%%%%%%%%%%%%%%%%%%%%%%%%%%%%%%%%%%%%%%%%%%%%%%%%%%%%%%%%%%%%%%%%%%%%%%%%%%%%%%%%%%%%%%%%%%%%%%%%%%%%%%%%%%%%%%%%%%%%%%%%%%%%%%%%%%%%%%%%%%%%%%%

\subsection{Determining the upper mass limit of a hypothetical perturber}

% Comments: Do not delete
% Written by Tobias C. Hinse
% The following text was largely taken from TEMPIV paper as the underlying methodology follows exactly that paper (and other papers).
% (This section is meant to be placed somewhere near the end of the manuscript).

Having transit timing measurements gives an opportunity to place constraints on the upper mass limit of a hypothetical planetary companion. The amplitude of the timing residuals is directly proportional to the mass of the perturber for a given orbital distance. The root-mean-square (RMS) statistic of the timing residuals offers a reasonable estimate of the amplitude which enables an estimate of the upper mass limit of a putative perturbing body. Any TTV effect is amplified for orbital configurations involving mean-motion resonances \citep{Agol+05mn,HolmanMurray05sci,NesvornyMorbidelli08apj} and should allow the detection of a low-mass planetary perturbing body. A larger perturbation implies a larger RMS scatter around the nominal ephemeris.

The applied method followed the technique described in \citet{Wang+17aj,Wang+18pasp,Wang+18aj2}. The calculation of an upper mass limit was performed numerically via direct orbit integrations of the equations of motion. For this task, we modified the {\sc fortran}-based {\sc microfarm}\footnote{\url{https://bitbucket.org/chdianthus/microfarm/src}} package \citep{Gozdziewski03aa,Gozdziewski++08mn} which relies on {\sc OpenMPI}\footnote{\url{https://www.open-mpi.org}} to spawn several parallel tasks. The package's main purpose is the numerical computation of the Mean Exponential Growth factor of Nearby Orbits \citep[MEGNO][]{CincottaSimo00aas,Gozdziewski+01aa,Cincotta++03phyd} over a grid of initial conditions of orbital parameters. In the present work we followed a direct brute-force method for the calculation of the RMS statistic.

\begin{figure*}
%% Toby's internal figure note (DO NOT DELETE!!):
%% Project path: /home/tobiash/simulations/JKT_HATPS18/TTVMEGNOMAPS/HATPS18
%% EPS file imported to GIMP, resolution 100, anti-aliasing for both graphics and text, rotated, cropped
%% (removing white border), added white alpha channel then exported as a PDF.
\includegraphics[width=1.02\columnwidth]{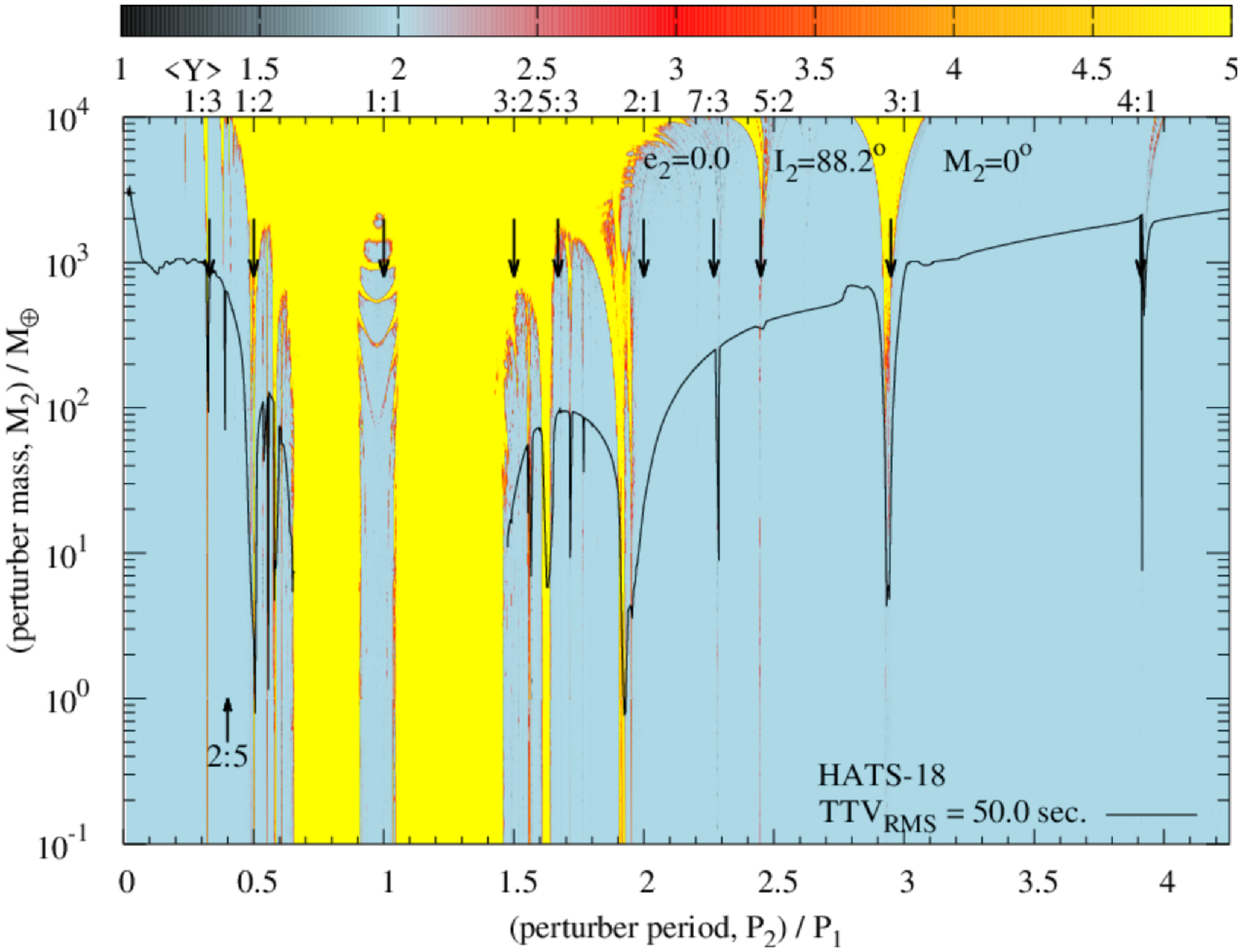}
\includegraphics[width=1.02\columnwidth]{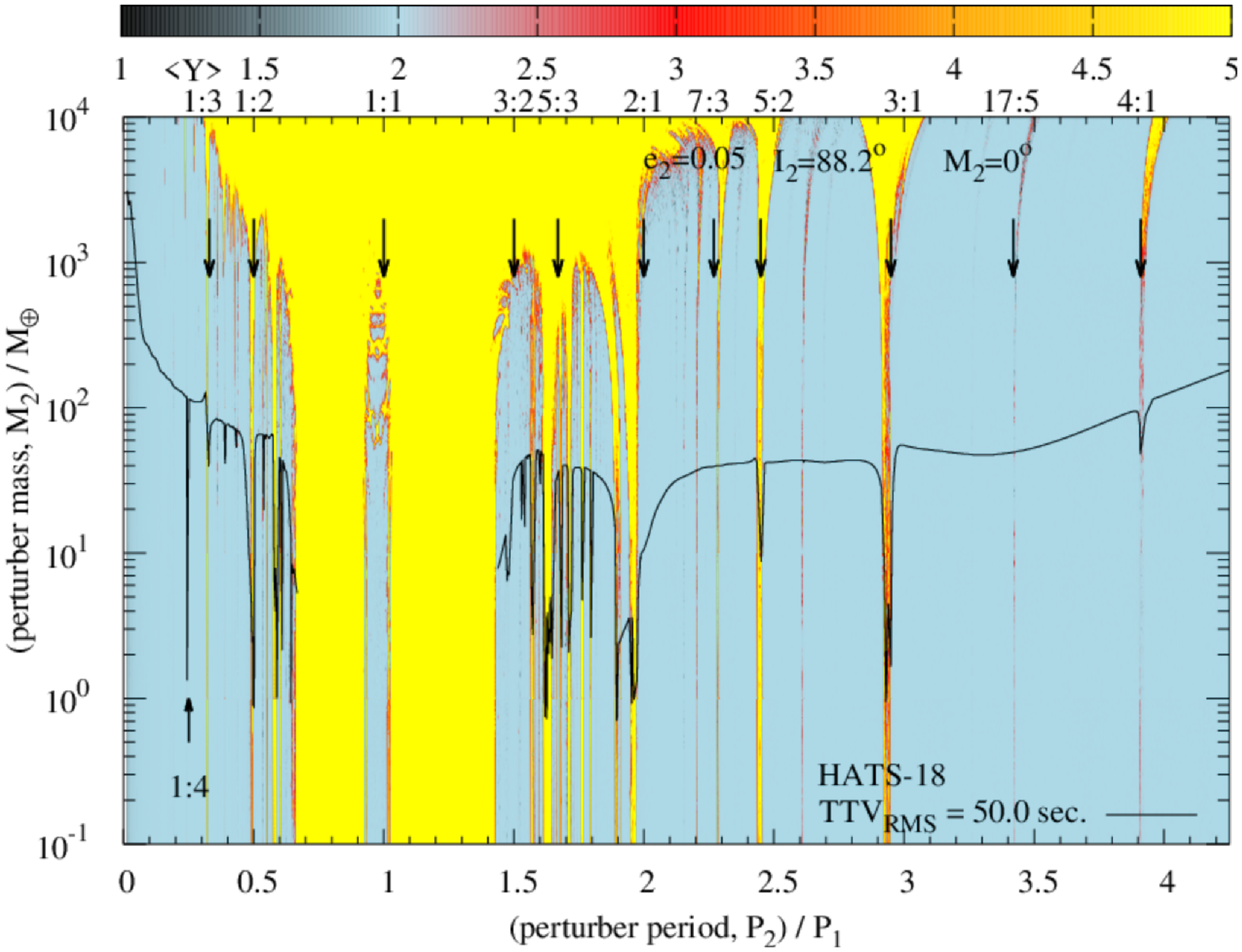}
\includegraphics[width=1.02\columnwidth]{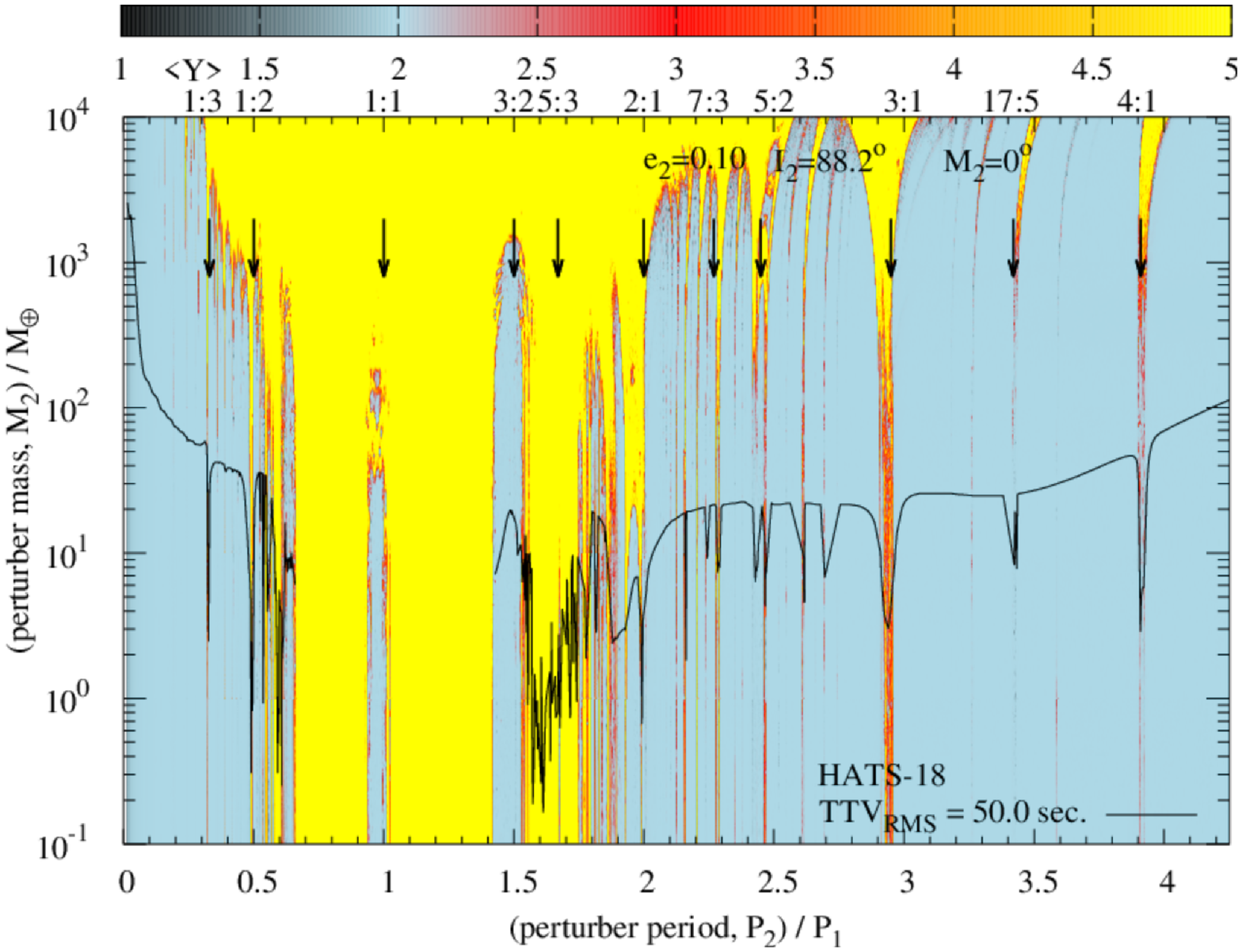}
\includegraphics[width=1.02\columnwidth]{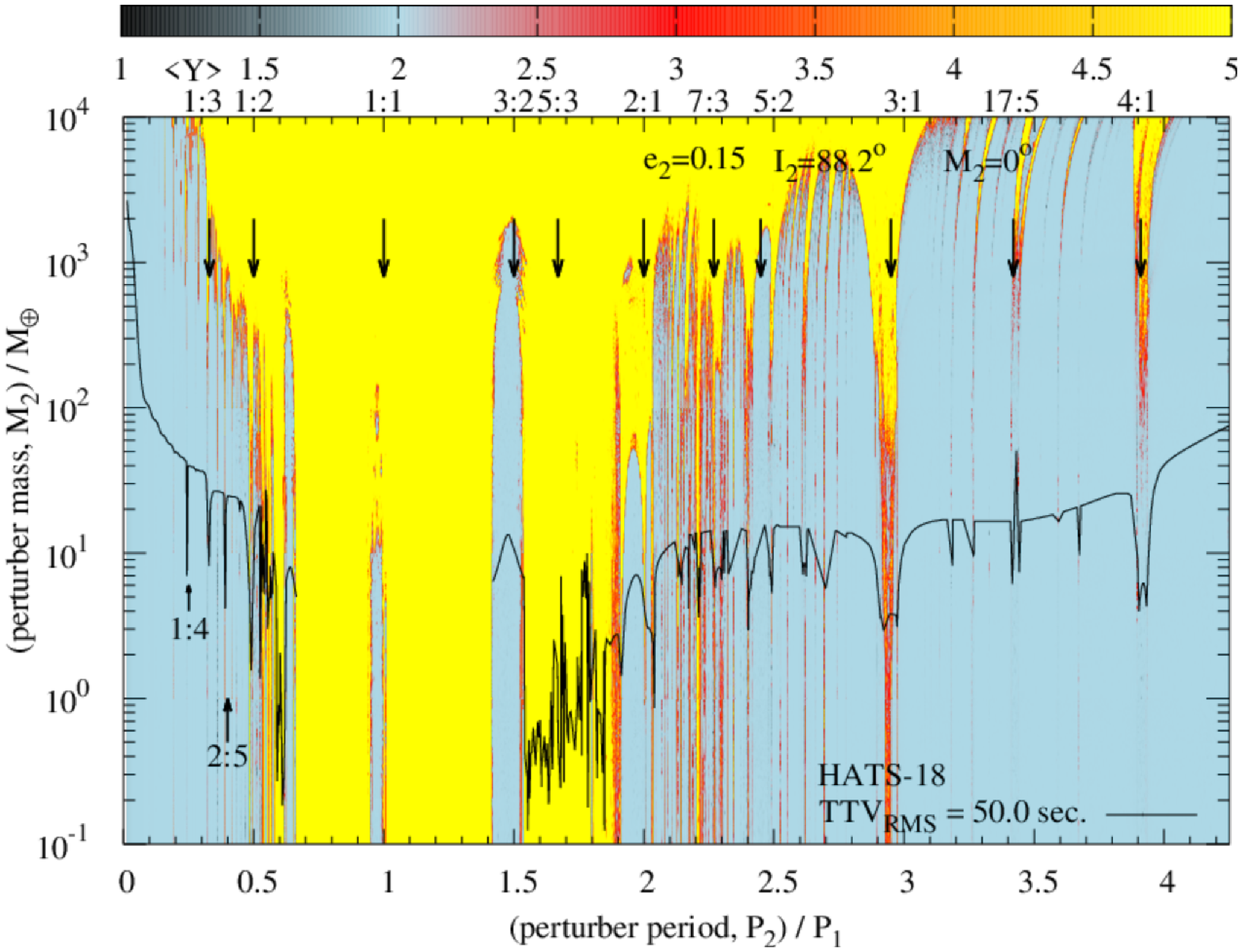}
\caption{Dynamical (heat) maps based on MEGNO and upper mass limit (solid line) of a putative perturbing
body for an observed TTV RMS statistic of 50\,s as determined in the present study. We considered four
different initial eccentricities for the perturbing planet.} \label{fig:megno_hats18}
\end{figure*}

We integrated the orbit of HATS-18 within the framework of the general three-body problem. The mid-transit time was calculated iteratively to a high precision from a series of back-and-forth integrations once a transit of HATS-18\,b was detected. The best-fit radii of both the planet and the host star were accounted for. We then calculated an analytic least-squares regression to the time-series of transit numbers and mid-transit times to determine a best-fitting linear ephemeris with an associated RMS statistic for the TTVs. The RMS statistic was based on a twenty-year integration corresponding to 8711 transits by HATS-18\,b. This procedure was then applied to a grid of masses and semimajor axes of the perturbing planet while fixing all the other orbital parameters. In this study, we chose to start the putative perturbing planet on a circular orbit that is coplanar with HATS-18\,b. This implies that $\Omega_2=0^{\circ}$ and $\omega_2=0^{\circ}$ for the perturber and $\Omega_1 = 0^{\circ}$ for the transiting planet. This setting provides the most conservative estimate of the upper mass limit of a possible perturber \citep{Bean09aa,Fukui+11pasj,Hoyer+11apj,Hoyer++12apj}. We refer the interested reader to \citet{Wang+18pasp}, who studied the effects of TTVs on varying initial orbital parameters of the perturbing body.

In order to calculate the location of mean-motion resonances, we used the same code to calculate MEGNO on the same parameter grid. However, this time we integrated each initial grid point for 1000 years, which was found to be long enough to allow detection of the location of weak chaotic high-order mean-motion resonances.

Briefly, the MEGNO factor $\langle Y(t) \rangle$ quantitatively measures the degree of stochastic behaviour of a nonlinear dynamical system and has proven useful in the detection of chaotic resonances \citep{Gozdziewski+01aa,Hinse+10mn}. In addition to the Newtonian equations of motion, the associated variational equations are solved simultaneously at each integration time step. The {\sc microfarm} package implements the {\sc odex}\footnote{\url{https://www.unige.ch/\~hairer/prog/nonstiff/odex.f}} extrapolation algorithm to numerically solve the system of first-order differential equations. This algorithm is slow but robust even for high-eccentricity orbits as well as close encounters. When presenting results (Fig.~\ref{fig:megno_hats18}) we always plot the time-averaged MEGNO that is utilized to quantitatively differentiate between quasi-periodic and chaotic dynamics. We refer to \cite{Hinse+10mn} for a short review of the essential properties of MEGNO.

In a dynamical system that evolves quasi-periodically in time the quantity $\langle Y \rangle$ will asymptotically approach 2.0 for $t \rightarrow \infty$. In that case, the orbital elements associated with that orbit are often bounded. In the case of a chaotic time evolution the quantity $\langle Y\rangle$ diverges away from 2.0, with orbital parameters exhibiting erratic temporal excursions. For quasi-periodic orbits, we typically find $|\langle Y \rangle - 2.0|$ to be less than 0.001 at the end of each integration.

Results are shown in Fig.~\ref{fig:megno_hats18} as MEGNO heat maps considering four different initial values in eccentricity for the putative perturbing planet. In each case, we find the usual instability region located in the proximity of the transiting planet (1:1 resonance) with MEGNO colour-coded as yellow (corresponding to $\langle Y\rangle > 5$). The extent of these regions coincides with the results presented in \citet{BarnesGreenberg06apj}. The locations of mean-motion resonances are indicated by arrows in each map.

We find that a perturbing body (initial circular orbit) of upper mass limit of around 1--$10$\,M$_{\oplus}$ will cause an RMS of the measured 52\,s when located in the $P_2/P_1=$ 2:1, 7:3, 5:2, 3:1 and 4:1 exterior resonance with HATS-18\,b. For the 1:3 and 1:2 interior resonance, perturber upper mass limits of $100$\,M$_{\oplus}$ and $1$\,M$_{\oplus}$, respectively, could also cause an RMS mid-transit time scatter of 52\,s. However, although we determined upper mass limits, in Section\,\ref{sec:ttv} we found no compelling evidence for any periodicities that can be attributed to an additional (non-transiting) planet within the HATS-18 system.

%%%%%%%%%%%%%%%%%%%%%%%%%%%%%%%%%%%%%%%%%%%%%%%%%%%%%%%%%%%%%%%%%%%%%%%%%%%%%%%%%%%%%%%%%%%%%%%%%%%%%%%%%%%%%%%%%%%%%%%%%%%%%%%%%%%%%%%%%%%%%%%%%%%%%

\section{Summary and conclusions}

HATS-18 consists of a massive planet (2\Mjup) on a very short-period orbit (0.838\,d) around a cool star (5600\,K) with a convective envelope. It is a promising candidate for the detection of orbital decay due to tidal effects \citep{Penev+16aj}, particularly as it is at the cusp of experiencing enhanced tidal dissipation due to internal gravity waves \citep{Barker20mn,MaFuller21apj}. We have obtained light curves of 20 transits from the 1.5\,m Danish telescope and nine transits from the Jongen telescope. We measured 27 times of mid-transit from these data, plus two timings from the TESS light curves of HATS-18 in sectors 10 and 36. Together with three published measurements, this gives a total of 32 transit timings available for the system.

The transit timings were fitted with several ephemerides in three functional forms (linear, quadratic and cubic), with the finding that the linear ephemeris represents them best. This equates to a non-detection of orbital decay which can be used to put an upper limit on the tidal quality factor of the star, $Q_\star^{\,\prime} = 10^{5.11 \pm 0.04}$. This constraint is similar to our theoretical predictions in Section\,\ref{sec:barker} if wave breaking occurs (or gravity waves are otherwise fully damped). This system therefore remains a very interesting one for follow-up studies with further observations having a strong potential to test tidal theory. One further sector of TESS data will (hopefully) become available in mid-2023, and additional ground-based timing measurements will help to strengthen the detection limit for orbital decay.

\begin{figure}
\includegraphics[width=\columnwidth,angle=0]{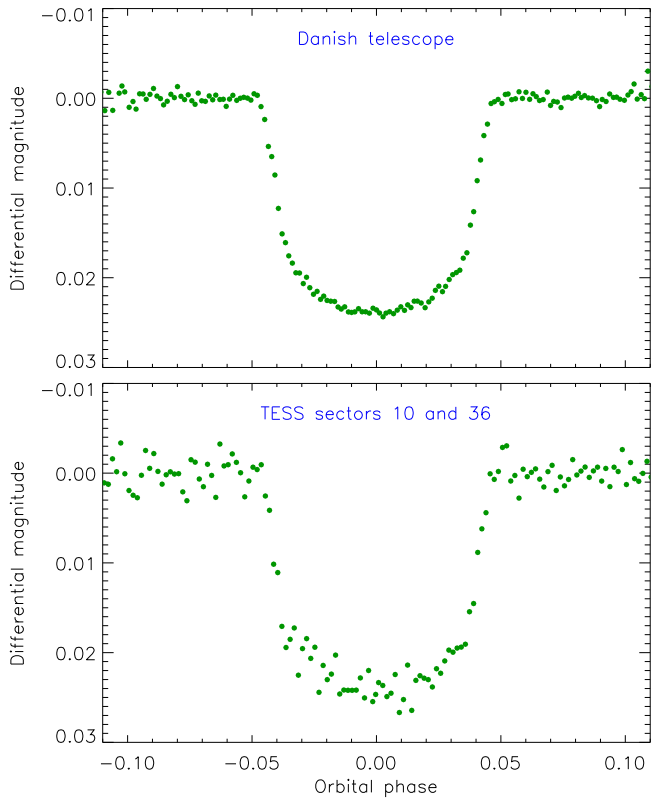}
\caption{\label{fig:compare} Transit observations of HATS-18 from the Danish
telescope (upper panel) and TESS (lower panel). Each dataset has beeen converted
into orbital phase then combined into bins of size 120\,s.} \end{figure}

We have also obtained revised measurements of the physical properties of the HATS-18 system which are in agreement with and improve on previous results. This part of the analysis was based on the data from the Danish telescope only, due to the higher precision of the observations. This prompted us to make a comparison between the light curves from the Danish telescope and TESS. The Danish telescope has an aperture of diameter 154\,cm and was used to observe 20 transits from the ground. TESS has an aperture of diameter 10.5\,cm and observed 52 transits over two sectors from space. We normalised each transit to zero differential magnitude, converted them into orbital phase, and binned them into bins of duration 120\,s. The results are shown in Fig.\,\ref{fig:compare}. It is clear that the ground-based observations are greatly superior to the space-based ones for a star of this relatively faint magnitude ($V=14.1$ and $i=13.8$; \citealt{Henden+12javso}). This is due to the much larger aperture, which lowers photon noise, and finer pixel scale, which lowers noise from the sky background even when the telescope is operated out of focus.

%%%%%%%%%%%%%%%%%%%%%%%%%%%%%%%%%%%%%%%%%%%%%%%%%%%%%%%%%%%%%%%%%%%%%%%%%%%%%%%%%%%%%%%%%%%%%%%%%%%%%%%%%%%%%%%%%%%%%%%%%%%%%%%%%%%%%%%%%%%%%%%%%%%%%

\section*{Acknowledgements}

We thank Dr.\ Kaloyan Penev and Dr.\ Joel Hartman for their help in tracking down the issue with the LCOGT transit timing.

AJB was supported by STFC grants ST/S000275/1 and ST/W000873/1.
UGJ acknowledges funding from the Novo Nordisk Foundation Interdisciplinary Synergy Programme grant no.\ NNF19OC0057374 and from the European Union H2020-MSCA-ITN-2019 under Grant no.\ 860470 (CHAMELEON).
PLP was partly funded by Programa de Iniciaci\'on en Investigaci\'on-Universidad de Antofagasta, INI-17-03.
NP's work was supported by Fundaca\~ao para a Ci\^encia e a Tecnologia (FCT) through the research grants UIDB/04434/2020 and UIDP/04434/2020.
This research has received funding from the Europlanet 2024 Research Infrastructure (RI) programme. The Europlanet 2024 RI provides free access to the world’s largest collection of planetary simulation and analysis facilities, data services and tools, a ground-based observational network and programme of community support activities. Europlanet 2024 RI has received funding from the European Union’s Horizon 2020 research and innovation programme under grant agreement No.\ 871149. This research received financial support from the National Research Foundation (NRF; No.\ 2019R1I1A1A01059609).

The following internet-based resources were used in research for this paper: the ESO Digitized Sky Survey; the NASA Astrophysics Data System; the SIMBAD database operated at CDS, Strasbourg, France; and the ar$\chi$iv scientific paper preprint service operated by Cornell University.

\section*{Data availability}

The light curves obtained with the Danish and Jongen telescopes will be made available at the Centre de Donn\'ees astronomiques de Strasbourg (CDS) at \texttt{http://cdsweb.u-strasbg.fr/}.
The TESS data used in this article are available in the MAST archive (\texttt{https://mast.stsci.edu/portal/Mashup/Clients /Mast/Portal.html}).

%%%%%%%%%%%%%%%%%%%%%%%%%%%%%%%%%%%%%%%%%%%%%%%%%%%%%%%%%%%%%%%%%%%%%%%%%%%%%%%%%%%%%%%%%%%%%%%%%%%%%%%%%%%%%%%%%%%%%%%%%%%%%%%%%%%%%%%%%%%%%%%%%%%%%

\bibliographystyle{mnras}
% \bibliography{jkt}

%%%%%%%%%%%%%%%%%%%%%%%%%%%%%%%%%%%%%%%%%%%%%%%%%%%%%%%%%%%%%%%%%%%%%%%%%%%%%%%%%%%%%%%%%%%%%%%%%%%%%%%%%%%%%%%%%%%%%%%%%%%%%%%%%%%%%%%%%%%%%%%%%%%%%
\end{document}